\documentclass[trackchanges,twocolumn]{aastex702}
\usepackage[utf8]{inputenc}

\newcommand{\ixpe}{\emph{IXPE}}
\newcommand{\xmm}{\emph{XMM-Newton}}
\newcommand{\nustar}{\emph{NuSTAR}}
\newcommand{\ep}{\emph{EP}}
\newcommand{\source}{EP J223759.5+531421}

\begin{document}

\title{\ixpe\ observation of the new magnetar source EP J223759.5+531421}

\author[0000-0003-3259-7801]{Roberto Taverna}
\affiliation{Department of Physics and Astronomy, University of Padova, via Marzolo 8, I-35131 Padova, Italy}
\email{taverna@pd.infn.it}

\author[0009-0001-4644-194X]{Lorenzo Marra}
\affiliation{INAF - Istituto di Astrofisica e Planetologia Spaziali, Via del Fosso del Cavaliere 100, I-00133, Roma}
\email{lorenzo.marra@inaf.it}

\author[0000-0003-3259-7801]{Roberto Turolla}
\affiliation{Department of Physics and Astronomy, University of Padova, via Marzolo 8, I-35131 Padova, Italy}
\affiliation{Mullard Space Science Laboratory, University College London, Holmbury St Mary, Dorking, Surrey RH5 6NT, UK}
\email{turolla@pd.infn.it}

\author[0000-0002-5004-3573]{Ruth M.E. Kelly}
\affiliation{Department of Physics and Astronomy, University of Padova, via Marzolo 8, I-35131 Padova, Italy}
\affiliation{Mullard Space Science Laboratory, University College London, Holmbury St Mary, Dorking, Surrey RH5 6NT, UK}
\email{rmek@mssl.ucl.ac.uk}

\author[0000-0001-5326-880X]{Silvia Zane}
\affiliation{Mullard Space Science Laboratory, University College London, Holmbury St Mary, Dorking, Surrey RH5 6NT, UK}
\email{sz@mssl.ucl.ac.uk}

\author[0000-0001-8785-5922]{Alice Borghese}
\affiliation{European Space Agency (ESA), European Space Astronomy Centre (ESAC), Camino Bajo del Castillo s/n, 28692 Villanueva de la Ca\~{n}ada, Madrid, Spain}
\email{alice.borghese@gmail.com}

\author[0000-0001-7611-1581]{Francesco Coti Zelati}
\affiliation{Institute of Space Sciences (ICE), CSIC, Campus UAB, Carrer de Can Magrans s/n, E-08193, Barcelona, Spain}
\affiliation{Institut d'Estudis Espacials de Catalunya (IEEC), Carrer Gran Capit\'a 2-4, E-08034, Barcelona, Spain}
\email{cotizelati@ice.csic.es}

\author[0000-0003-4849-5092]{Paolo Esposito}
\affiliation{University School for Advanced Studies IUSS Pavia, Palazzo del Broletto, Piazza della Vittoria 15, I-27100 Pavia, Italy}
\email{paolo.esposito@iusspavia.it}

\author[0000-0001-5480-6438]{Gian Luca Israel}
\affiliation{INAF - Osservatorio Astronomico di Roma, via Frascati 33, I-00078 Monteporzio Catone, Italy}
\email{gianluca.israel@inaf.it}

\author[0000-0003-2177-6388]{Nanda Rea}
\affiliation{Institute of Space Sciences (ICE), CSIC, Campus UAB, Carrer de Can Magrans s/n, E-08193, Barcelona, Spain}
\affiliation{Institut d'Estudis Espacials de Catalunya (IEEC), Carrer Gran Capit\'a 2-4, E-08034, Barcelona, Spain}
\email{rea@ice.csic.es}

\author[0000-0002-7680-2056]{Haonan Yang}
\affiliation{National Astronomical Observatories, Chinese Academy of Sciences, 20A Datun Road, Beijing 100101, China}
\email{hnyang@bao.ac.cn}

\author[0009-0009-1721-3663]{Yilong Wang}
\affiliation{Institute of Space Sciences (ICE), CSIC, Campus UAB, Carrer de Can Magrans s/n, E-08193, Barcelona, Spain}
\affiliation{Institut d'Estudis Espacials de Catalunya (IEEC), Carrer Gran Capit\'a 2-4, E-08034, Barcelona, Spain}
\email{ywang@ice.csic.es}


\begin{abstract}
We report on the detection of polarized X-ray emission from the new Galactic magnetar \source, discovered by the Einstein Probe Wide-field X-ray Telescope on 2026 June 28. The Imaging X-ray Polarimetry Explorer (\ixpe) follow-up observation started on 2026 July 6 and detected the source at a (absorbed) flux of $\approx 4.5\times 10^{-11} \, \mathrm{ erg\,cm^{-2}\,s}^{-1}$ ($2$--$7.5\, \mathrm{keV}$ range), confirming the presence of two thermally emitting regions on the star surface, with temperatures $\approx 0.5$ and $\approx 1\, \mathrm{keV}$. A significant (at $> 4\sigma$ confidence level) phase- and energy-integrated polarization degree of $\approx 7\%$ was detected with the polarization angle $\approx -2^\circ$ East of the celestial North. The two thermal components exhibit quite different polarization properties. The difference is larger in the phase-folded data, with the polarization degree reaching $\sim 63\%$ in the range $4$--$7.5\,\mathrm{keV}$, in correspondence with the secondary peak of the pulse profile, and never exceeding $\sim 30\%$ at lower energies. The polarization angle continuously oscillates from $-90^\circ$ to $+90^\circ$ over one rotational cycle and is well fit by the rotating vector model, with an inclination of the line of sight and of the magnetic dipole axis relative to the star spin axis of $\approx 28^\circ$  and $\approx 109^\circ$, respectively. The data point to an emission geometry in which the warm thermal component could originate from two antipodal caps, possibly in a magnetically condensed state, while the hot one could come from a smaller region covered by an atmosphere.

\end{abstract}


\section{Introduction}
\setcounter{footnote}{0}

On 28 June 2026, the \emph{Einstein Probe} \citep[\ep;][]{2022hxga.book...86Y} detected a previously unreported transient X-ray source exhibiting the traits of a potential new magnetar in outburst: unabsorbed $0.5$--$10\,\mathrm{keV}$ flux $\sim 10^{-10}\,\mathrm{erg\,cm^{-2}\,s^{-1}}$, an absorbed power-law spectrum with photon index $\sim 2$ and a clear periodic modulation, with  $P\sim 3\,\mathrm{s}$ \cite[][]{2026ATel17859....1Y}, later realized to be the first harmonic of a $6\,\mathrm{s}$ spin signal \citep{2026ATel17870....1R}. 

Follow-up observations with \ep\ and \nustar\ confirmed the source, named \source, as a magnetar candidate \citep{2026ATel17870....1R}. The magnetar nature of \source\ has been further strengthened by the detection of seven short ($20$--$40\, \mathrm{ms}$) bursts with the ART-XC telescope on board the SRG observatory \citep{2026ATel17908....1M}, as well as an additional $40\, \mathrm{ms}$ burst with ECLAIRs on board SVOM \citep{2026GCN.45270....1F}. The source spectrum in the \emph{NuSTAR} observation was well represented by the superposition of two thermal components, with $kT_\mathrm w\sim 0.4 \ \mathrm{keV}$ and $kT_\mathrm h\sim 1 \ \mathrm{keV}$, and a 
power-law tail with spectral index $\Gamma\sim 1.4$ extending to $\sim 30 \ \mathrm{keV}$. The neutral hydrogen column density was constrained to be $N_\mathrm{H}\sim 6.2\times 10^{21}\ \mathrm{cm}^{-2}$, with unabsorbed flux $F_\mathrm{X}\sim 1.6\times 10^{-10}\ \mathrm{erg\,cm^{-2}\,s^{-1}}$ in the range $0.5$--$10\ \mathrm{keV}$ \cite[][]{2026ATel17859....1Y}. 
Early searches at radio, optical, and near-infrared wavelengths did not reveal a counterpart \citep{atelradio1,atelradio2,ateloptical1,2026ATel17877....1Z}.  

Further monitoring with \ep, \xmm, \emph{SVOM} and \nustar\ between  June 29 and July 18 2026 provided a more accurate source position at $\mathrm{RA} = 22^\mathrm{h}37^\mathrm{m}59^\mathrm{s}.4\,, \mathrm{DEC}= +53^\circ14'22''.7$ (J2000.0) and a preliminary phase-coherent timing solution with $P=5.9958723(4)\, \mathrm s$ and $\dot P= 5(1)\times 10^{-12}\, \mathrm{s/s}$ \cite[][]{cotiatel,2026ATelgotz}. The implied surface dipolar field at the equator is $B\sim  2\times10^{14}\, \mathrm G$, confidently qualifying \source\ as a magnetar. A detailed analysis of these and the forthcoming observations will be reported in a separate paper \citep{Rea2026subm}. 

The \ixpe\ observatory \citep{2022JATIS...8b6002W} opened a new window in the X-ray sky, finally making it possible to measure the X-ray polarization in magnetar sources \citep[see][for a summary of \ixpe\ results]{2024Galax..12....6T}. In this {\emph{Letter}}, we report on the first \ixpe\ observation of \source\ carried out in response to a ToO proposal (PI: R. Taverna).  

\section{Observations and data reduction} \label{sec:datareduction}
\begin{figure*}[]
\includegraphics[width=1.\textwidth]{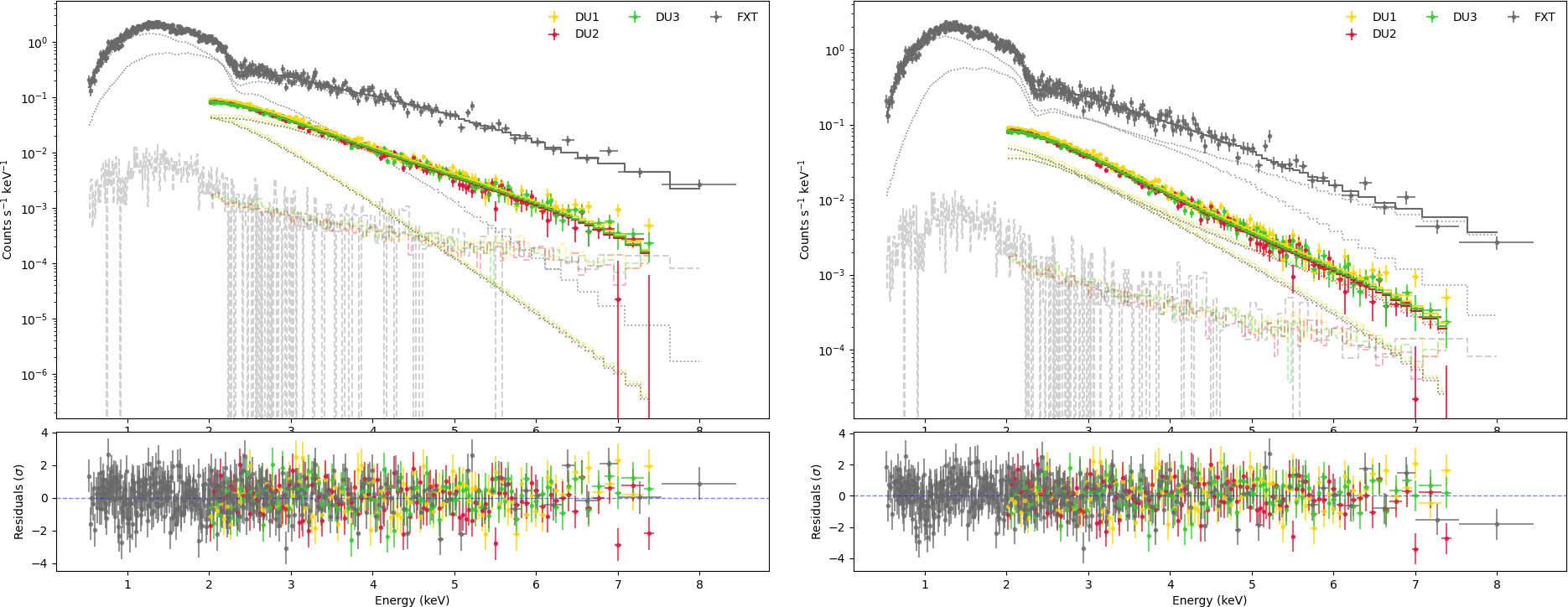}
\caption{Phase-integrated count spectra (dots with $1\sigma$ error bars) collected by \ixpe\ DU 1 (yellow), 2 (red) and 3 (green) and by {\it EP-FXT} (gray), fitted with a two-blackbody model (left panel) and a blackbody+power-law model (right panel). The background for each detector and the single model components are also shown by dashed and dotted histograms, respectively. Fit residuals for the two fits are also reported in the bottom panels. The best-fit parameters are listed in Table \ref{tab:specpa}.
\label{fig:spectrumphaseint}}
\end{figure*}

\subsection{IXPE}
The \ixpe\ observation of \source\ started on 2026 July 6 09:18:02.102 UTC and ended on 2026 July 8 16:53:23.425 UTC, for a total elapsed time of $\approx200\,\mathrm{ks}$. We retrieved level 1 and level 2 data files from the HEASARC \ixpe\ public archive\footnote{\url{https://heasarc.gsfc.nasa.gov/docs/ixpe/archive}} and referred the event arrival times to the Solar System barycenter using the ftool \texttt{barycorr} with JPL planetary ephemeris DE440 and the coordinates $\mathrm{RA} = 22^\circ37'59''.4\,, \mathrm{DEC}= +53^\circ14'22''.7$ \citep{cotiatel}. Background rejection was performed following the method presented in \citet{2023AJ....165..143D}, with the appropriate modifications to include the corrections for \ixpe\ detector unit (DU) 2, following the failure experienced in 2025 \cite[see e.g.][]{2026arXiv260403366D}. We checked that  no solar flares occurred during the observation of \source\ by examining the background lightcurve (so no excision of time intervals was necessary). 
The source and background counts were extracted from the processed photon lists in  a circle with radius $r_\mathrm{src}$ and an annulus with inner and outer radii $r_\mathrm{bkg}^\mathrm{inn}$ and $r_\mathrm{bkg}^\mathrm{out}$, respectively, both centered on the source position. By maximizing the signal-to-noise ratio as discussed in \citet{2004MNRAS.351..161P}, we chose $r_\mathrm{src}=80''$, $r_\mathrm{bkg}^\mathrm{inn}=120''$ and $r_\mathrm{bkg}^\mathrm{out}=240''$. A total of $\approx 60\,000$ net source counts were collected over the three \ixpe\ DUs in the entire $2$--$8\,\mathrm{keV}$ band. 

\subsection{EP-FXT}
We analysed two observations of \source\ performed by the Follow-up X-ray Telescope (FXT) aboard \ep\ during the \ixpe\ observing window (ObsIDs 08500000643 and 08500000644), totalling $10.3\,\mathrm{ks}$ of cleaned exposure. We used only FXT-A data, acquired in partial-window imaging mode, because its spectral calibration is better characterized than that of the corresponding FXT-B timing-mode data. Each dataset was processed with FXTDAS v1.30. Source and background events were extracted from a $60\arcsec$-radius circle and a $120\arcsec$--$240\arcsec$ annulus, respectively. Event times were referred to the Solar System barycenter using the DE440 ephemeris and the above-mentioned coordinates, and source, background spectra and response files were generated. For the phase-averaged analysis, the spectra from the two observations were combined. For the phase-resolved analysis, the events were folded using the timing solution described in Section~\ref{sec:timing} and divided into the same seven equally spaced phase intervals adopted for the \ixpe\ analysis. Spectra were extracted separately for each observation and phase interval and combined across observations within each interval. The resulting spectra were grouped over the $0.5$--$10\,\mathrm{keV}$ energy range to a minimum of 25 background-subtracted counts per bin.

\section{Timing analysis} \label{sec:timing}

We extracted the background-subtracted power spectrum from the \ixpe\ dataset and used {\sc HENdrics}-{\sc stingray} software to search for periodic signals \cite[see][]{2015ascl.soft02021B,2018ascl.soft05019B}. A prominent peak was found at frequency $\nu\approx 0.166\,\mathrm{Hz}$, followed by its first two harmonics, in agreement with the preliminary solution reported by \citet{2026ATel17870....1R}. We performed a  $Z^2_\mathrm{n}$-search (with $n=3$) around the fundamental frequency by splitting the observation into $128$ time intervals. The  corresponding times of arrival (TOAs) were calculated using the {\tt HENphaseogram} tool. Finally, we fitted the TOAs with a linear spin-down model with {\sc pint} \cite[][]{2021ApJ...911...45L}, obtaining a best-fit spin frequency $\nu=0.16678132\pm1.8\times10^{-7}\,\mathrm{Hz}$ and frequency derivative $\dot{\nu}=(-2.0\pm40.9)\times10^{-13}\,\mathrm{Hz\,s}^{-1}$ (spin period $P=5.9958751\pm6.6\times10^{-6}\,\mathrm{s}$ and spin-down rate $\dot{P}=(7.3\pm146.9)\times10^{-12}\,\mathrm{s\,s}^{-1}$)\footnote{Here and hereafter errors are given at $1\sigma$ confidence level, unless explicitly stated otherwise.}. Our solution is in agreement with that reported by \citet{cotiatel} within the errors, although our relatively short baseline leaves  $\dot P$  completely unconstrained.
\begin{table*}[!t]
\tabletypesize{\scriptsize}
\begin{center}
\caption{Best-fit spectral parameters of the joint  phase-averaged \ixpe-{\it EP} data. \label{tab:specpa}}
\begingroup
\setlength{\tabcolsep}{18.0pt}
\renewcommand{\arraystretch}{1.6}
\begin{tabular}{c | c | c}
\hline\hline
\  & \texttt{constant}$\times$\texttt{tbabs}$\times$\texttt{(bbodyrad}$+$\texttt{bbodyrad)} & \texttt{constant}$\times$\texttt{tbabs}$\times$\texttt{(bbodyrad}$+$\texttt{powerlaw)} \\
\hline
$\mathrm{const}_{\mathrm{DU}_1}$ $^\mathrm{(a)}$& $1$ $^\mathrm{(b)}$ & $1$ $^\mathrm{(b)}$ \\
$N_\mathrm{H}$ ($10^{22}\,\mathrm{cm}^{-2}$) & $0.209^{+0.016}_{-0.016}$ & $0.546^{+0.026}_{-0.026}$ \\
$kT_1$ ($\mathrm{keV}$) & $0.449^{+0.015}_{-0.014}$ & $0.763^{+0.018}_{-0.018}$ \\
$R_{\mathrm{BB}_1}\, (\mathrm{km}$) $^\mathrm{(c)}$ & $2.616^{+0.146}_{-0.127}$ & $0.911^{+0.045}_{-0.041}$ \\
$kT_2$ ($\mathrm{keV}$) & $1.048^{+0.025}_{-0.023}$ & --- \\
$R_{\mathrm{BB}_2}\, (\mathrm{km})$ $^\mathrm{(c)}$ & $0.635^{+0.037}_{-0.035}$ & --- \\
$\Gamma_\mathrm{PL}$ & --- & $2.134^{+0.061}_{-0.059}$ \\
$\mathrm{norm}_\mathrm{PL}$ $^\mathrm{(d)}$ & --- & $0.015^{+0.001}_{-0.001}$ \\
$\mathrm{const}_{\mathrm{DU}_2}$ $^\mathrm{(a)}$ & $0.980^{+0.010}_{-0.010}$ & $0.980^{+0.010}_{-0.010}$ \\
$\mathrm{const}_{\mathrm{DU}_3}$ $^\mathrm{(a)}$& $0.981^{+0.010}_{-0.010}$ & $0.981^{+0.010}_{-0.010}$ \\
$\mathrm{const}_\mathrm{EP}$ $^\mathrm{(a)}$& $0.953^{+0.012}_{-0.012}$ & $0.961^{+0.012}_{-0.012}$\\
$F_\mathrm{unabs}^{2.0-7.5}$ $^\mathrm{(e)}$ & $4.584^{+0.016}_{-0.016}$ & $4.749^{+0.016}_{-0.016}$\\
$F_\mathrm{obs}^{2.0-7.5}$ $^\mathrm{(f)}$ & $4.501^{+0.015}_{-0.016}$ & $4.525^{+0.016}_{-0.016}$\\
$\chi^2/\mathrm{dof}$ & $674.27/694$ & $687.50/694$ \\
\hline\hline
\end{tabular}
    \endgroup
    \end{center}
    \tablecomments{ \\
    $\mathrm{(a)}$ -- Cross calibration factor of the three \ixpe\ DUs and {\it EP-FXT}, taken to be unity for \ixpe\ DU 1.\\
    $\mathrm{(b)}$ -- Frozen parameters.\\
    $\mathrm{(c)}$ -- Emission radius extracted from the {\tt bbodyrad} parameter $\mathrm{norm}$ assuming a distance of $3.3\,\mathrm{kpc}$. \\
    $\mathrm{(d)}$ --  Normalization of the {\tt powerlaw} component in units of photons $\mathrm{keV}^{-1}\,\mathrm{cm}^{-2}\,\mathrm{s}^{-1}$ at $1\,\mathrm{keV}$. \\
    $\mathrm{(e)}$ -- Unabsorbed $2.0$--$7.5\,\mathrm{keV}$ flux in units of $10^{-11}\,\mathrm{erg\,cm^{-2}\,s^{-1}}$. \\
    $\mathrm{(f)}$ -- Observed $2.0$--$7.5\,\mathrm{keV}$ flux in units of $10^{-11}\,\mathrm{erg\,cm^{-2}\,s^{-1}}$.}
\end{table*}

\section{Spectral analysis} \label{sec:spectral}

Spectral analysis of the \ixpe\ data was performed with the {\sc xspec} suite \cite[][]{1996ASPC..101...17A}, after generating the ancillary (ARFs) and modulation response  (MRFs) files from the most recent version\footnote{They were downloaded from the \ixpe\ calibration center database, \url{https://heasarc.gsfc.nasa.gov/docs/ixpe/caldb}.} of the \ixpe\ response functions (v20260610) using the  {\tt ixpecalcarf} tool \cite[][]{2025AAS...24526001C}. Data grouping was performed to ensure a signal-to-noise ratio $\gtrsim3$ in each energy bin. Since the source signal is background-dominated above $\sim 7.5\,\mathrm{keV}$, in the spectro-polarimetric analysis we limited the \ixpe\ dataset to the $2.0$--$7.5\,\mathrm{keV}$ range.

We fit jointly the simultaneous \ep\ (total exposure time of $\approx10\,\mathrm{ks}$) and \ixpe\ observations, allowing us to extend the energy range down to $0.5\,\mathrm{keV}$ and improve the statistical significance of our dataset.

A fit with a single-component, absorbed model for the combined \ixpe-{\it EP} dataset fails the $99\%$ goodness-of-fit criterion ($\chi^2/\mathrm{dof}=1423.86/696$ for a single blackbody, BB, and $1147.54/696$ for a single power-law, PL). We then attempted a fit with two-component models, either two blackbodies  (\texttt{bbodyrad}$+$\texttt{bbodyrad}) or a blackbody plus a power-law (\texttt{bbodyrad}$+$\texttt{powerlaw}). Interstellar absorption was taken into account using the \texttt{tbabs} model, with photoionization cross sections of \citet{1996ApJ...465..487V} and abundances of \citet{2000ApJ...542..914W}. 
A normalization constant was included in the spectral model to account for cross-calibration differences among the three \ixpe\ DUs and {\it EP-FXT}. 

Both BB+BB and BB+PL  fits were found to be statistically acceptable, the former providing a better fit to the data at the same number of degrees of freedom ($\chi^2/\mathrm{dof}=674.27/694$ vs. $\chi^2/\mathrm{dof}=687.50/694$, i.e., $\Delta\chi^2=13.23$). Moreover, the BB+PL model exhibits a systematic residual structure at low energies, where a noticeable oscillatory pattern is apparent (see Figure \ref{fig:spectrumphaseint}). We also find a significant increase in $N_\mathrm{H}$ as a PL component is included in the spectrum. Taken together, these considerations motivate our choice to adopt the purely thermal model. The resulting two-blackbody thermal decomposition is also consistent with the presence of two emitting regions already found in the previous analyses \cite[][]{2026ATel17870....1R}: a warm, more extended one with temperature $kT_\mathrm{w}\approx0.5\,\mathrm{keV}$, alongside a hot, smaller one, with temperature $kT_\mathrm{h}\approx1.1\,\mathrm{keV}$; the radiation radii are $R_\mathrm w\sim 2.6\, \mathrm{km}$ and $R_\mathrm h\sim 0.6\, \mathrm{km}$, respectively, assuming a distance of $3.3\, \mathrm{kpc}$\footnote{The distance $D$ of \source\ is still unknown but the value of $N_\mathrm{H}$ argues in favor of $D\lesssim 5\, \mathrm{kpc}$ and its position on the sky makes an association with the Perseus spiral arm likely. 
The line-of-sight distance of $3.3\,\mathrm{kpc}$ corresponds to the local maximum of the Galactic hydrogen density profile within the Perseus arm \cite[see Appendix B in][]{Rea2026subm}.}. 
The phase-integrated spectral fits to the joint \ixpe-{\it EP} data are shown in Figure \ref{fig:spectrumphaseint}, with the corresponding best-fit parameters reported in Table \ref{tab:specpa}.

We then phase-folded the \ixpe\ data using the timing solution discussed in section \ref{sec:timing}, by dividing the rotational phase into seven equally-spaced  bins with the {\tt xpphase} and {\tt xpselect} tools of the {\sc ixpeobssim} suite \cite[][]{2022SoftX..1901194B}, accounting for the exact number of counts with the {\sc ftool} {\tt fcalc}. The {\it EP-FXT} data were phase-folded following the same timing solution, ensuring proper alignment of the initial phase with the \ixpe\ dataset. Finally, we applied the same procedure that was used for the phase-averaged spectral analysis to each phase bin. 

\begin{figure}[]
\includegraphics[width=0.47\textwidth]{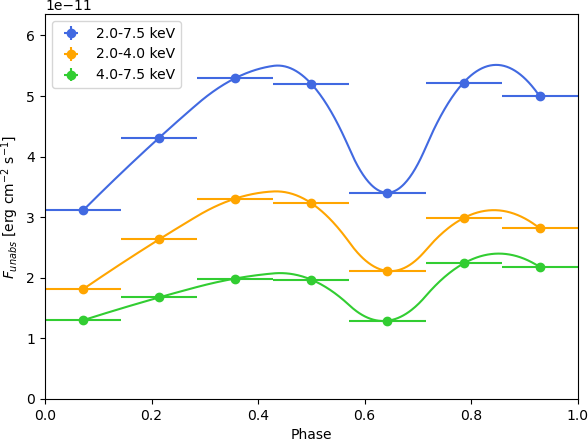}
\caption{Pulse profile of \source\ from the \ixpe-{\it EP} joint analysis in the $2.0$--$7.5\,\mathrm{keV}$ (cyan), $2.0$--$4.0\,\mathrm{keV}$ (orange) and $4.0$--$7.5\,\mathrm{keV}$ (green) energy bands. The error bars, at $1\sigma$ confidence level (see Table \ref{tab:pulseprof}), lie inside the symbols. The curves are splines connecting the data for a better visualization.
\label{fig:lightcurves}}
\end{figure}
Figure \ref{fig:lightcurves} reports the lightcurves in the energy bands $2.0$--$7.5$, $2.0$--$4.0$ and $4.0$--$7.5\,\mathrm{keV}$. The pulse profile is double-peaked and its overall shape appears fairly unchanged with energy, with a first broad peak between phases $0.0$ and $0.6$ and a second narrower peak between phases $0.6$ and $1.0$. Nevertheless, the relative amplitude of the two peaks shows a weak energy dependence, with the second peak slightly suppressed with respect to the first one at low energies, while the trend is reversed at high energies. A mild energy dependence is also observed in the behavior of the pulsed fraction, $\mathrm{PF}=(F_\mathrm{max}-F_\mathrm{min})/(F_\mathrm{max}+F_\mathrm{min})$, which stands at approximately $30\%$ in the $2.0$--$4.0\,\mathrm{keV}$ band and drops to about $27\%$ in the $4.0$--$7.5\,\mathrm{keV}$ range (across the entire $2.0$--$7.5\,\mathrm{keV}$ band, the overall pulsed fraction is around $26\%$).

\suppressfloats[t]
\begin{figure}[!t]
\includegraphics[width=0.47\textwidth]{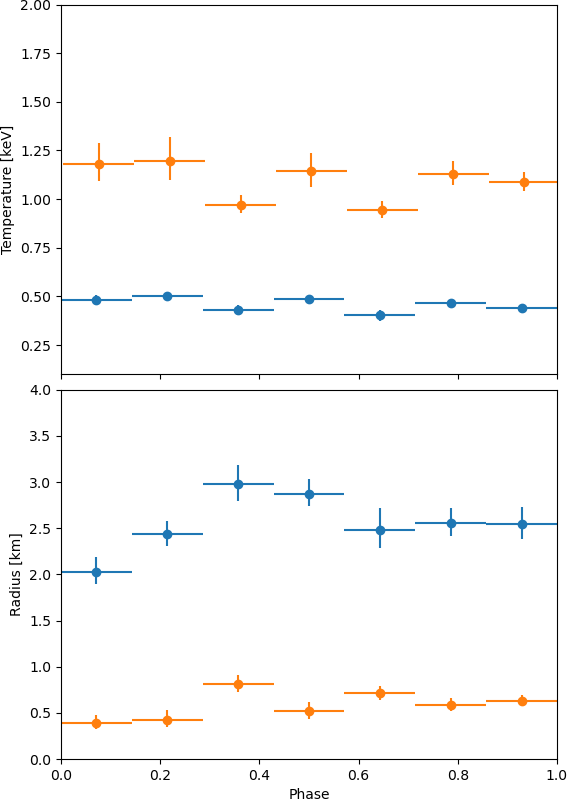}
\caption{Temperature (top) and emitting radius (bottom) of the warm (cyan) and hot (orange) blackbody spectral components as a function of the rotational phase. Error bars are given at $1\sigma$ confidence level (numerical values are reported in Table \ref{tab:specpr}).
\label{fig:param_spec_pr}}
\end{figure}
A two-blackbody decomposition also provides a good fit to the spectrum in each phase interval, with spectral parameters well constrained in all phase bins. The best fitting parameters are reported in Table \ref{tab:specpr},  while Figure \ref{fig:param_spec_pr} shows the variation  of the temperatures and the emission radii with the phase. A fit with a constant function yields $p$-values $\lesssim5\%$, indicating that the parameters of both BB components are inconsistent with being constant over the rotational phase. Nonetheless, a comparison of these trends with the double-peaked lightcurves shown in Figure \ref{fig:lightcurves} reveals only a marginal resemblance, more pronounced for the emitting radius of the warm BB component (for which a constant fit returned $p=8.1\times10^{-4}$). In particular, while both blackbody radii attain a local maximum in correspondence with the first peak (at phase $\approx 0.35$), the two blackbody temperatures exhibit rather a minimum at the same rotational phase. On the other hand, the mean values inferred from constant fits to the four parameter distributions ($\overline{kT}_\mathrm{w}=0.462\pm0.008\,\mathrm{keV}$, $\bar{R}_{\mathrm{BB}_\mathrm{w}}=2.561\pm0.058\,\mathrm{km}$, $\overline{kT}_\mathrm{h}=1.057\pm0.019\,\mathrm{keV}$ and $\bar{R}_{\mathrm{BB}_\mathrm{h}}=0.598\pm0.028\,\mathrm{km}$) are all consistent with the phase-averaged values reported in Table \ref{tab:specpa} within $1\sigma$ uncertainties.

\begin{table*}[ht!]
\tabletypesize{\scriptsize}
\begin{center}
\caption{Best-fit spectral parameters of the joint phase-resolved \ixpe-{\it EP} data. \label{tab:specpr}}
\begingroup
\renewcommand{\arraystretch}{1.4}
\begin{tabular}{c}
Spectral model: \texttt{constant}$\times$\texttt{tbabs}$\times$\texttt{(bbodyrad$+$bbodyrad)}$^\mathrm{(a)}$ \\
\end{tabular}
\setlength{\tabcolsep}{5.5pt}
\begin{tabular}{c c c c c c c c c}
\hline\hline
phase bin & $kT_\mathrm{w}$ ($\mathrm{keV}$) & $R_{\mathrm{BB_w}}$ ($\mathrm{km}$)$^\mathrm{(b)}$ & $kT_\mathrm{h}$ ($\mathrm{keV}$) & $R_{\mathrm{BB_h}}$ ($\mathrm{km}$)$^\mathrm{(b)}$ & $\mathrm{const_{DU_2}}$ & $\mathrm{const_{DU_3}}$ & $\mathrm{const_{EP}}$ & $\chi^2/\mathrm{dof}$ \\
\hline
$0.00$--$0.14$ & $0.483^{+0.026}_{-0.025}$ & $2.029^{+0.157}_{-0.133}$ & $1.183^{+0.108}_{-0.089}$ & $0.393^{+0.082}_{-0.062}$ & $0.955^{+0.032}_{-0.031}$ & $1.014^{+0.034}_{-0.033}$ & $0.912^{+0.037}_{-0.036}$ & $265.16/346$ \\
$0.14$--$0.29$ & $0.502^{+0.022}_{-0.023}$ & $2.434^{+0.148}_{-0.129}$ & $1.197^{+0.123}_{-0.098}$ & $0.426^{+0.105}_{-0.075}$ & $1.007^{+0.028}_{-0.027}$ & $0.965^{+0.027}_{-0.026}$ & $0.977^{+0.032}_{-0.031}$ & $329.31/399$ \\
$0.29$--$0.43$ & $0.430^{+0.024}_{-0.024}$ & $2.973^{+0.208}_{-0.177}$ & $0.971^{+0.050}_{-0.042}$ & $0.813^{+0.101}_{-0.090}$ & $0.955^{+0.024}_{-0.023}$ & $0.972^{+0.024}_{-0.024}$ & $0.974^{+0.028}_{-0.028}$ & $405.79/433$ \\
$0.43$--$0.57$ & $0.486^{+0.016}_{-0.014}$ & $2.872^{+0.156}_{-0.137}$ & $1.143^{+0.094}_{-0.078}$ & $0.519^{+0.103}_{-0.078}$ & $0.964^{+0.024}_{-0.024}$ & $0.967^{+0.024}_{-0.024}$ & $0.906^{+0.027}_{-0.026}$ & $367.60/362$ \\
$0.57$--$0.71$ & $0.402^{+0.028}_{-0.028}$ & $2.482^{+0.239}_{-0.195}$ & $0.944^{+0.049}_{-0.041}$ & $0.713^{+0.084}_{-0.077}$ & $0.981^{+0.031}_{-0.030}$ & $0.961^{+0.030}_{-0.030}$ & $0.965^{+0.035}_{-0.034}$ & $270.01/330$ \\
$0.71$--$0.86$ & $0.465^{+0.022}_{-0.022}$ & $2.554^{+0.165}_{-0.144}$ & $1.129^{+0.066}_{-0.057}$ & $0.582^{+0.077}_{-0.066}$ & $0.983^{+0.026}_{-0.025}$ & $1.015^{+0.026}_{-0.026}$ & $0.946^{+0.029}_{-0.028}$ & $343.75/419$ \\
$0.86$--$1.00$ & $0.438^{+0.023}_{-0.022}$ & $2.544^{+0.183}_{-0.158}$ & $1.088^{+0.052}_{-0.045}$ & $0.633^{+0.066}_{-0.059}$ & $1.017^{+0.027}_{-0.027}$ & $0.975^{+0.026}_{-0.026}$ & $0.966^{+0.031}_{-0.030}$ & $329.59/398$ \\
\hline\hline
\end{tabular}
    \endgroup
    \end{center}
    \tablecomments{\\
    (a) -- $N_\mathrm{H}$ is fixed at the value found in the phase-averaged spectral fit ($2.09\times10^{21}\,\mathrm{cm}^{-2}$, see Table \ref{tab:specpa}) and the cross-calibration constant for \ixpe\ DU 1 is fixed to unity. \\
    (b) -- Emission radius extracted from the \texttt{bbodyrad} parameter $\mathrm{norm}$ assuming a distance of $3.3\,\mathrm{kpc}$.}
\end{table*}

\begin{table*}[ht!]
\tabletypesize{\scriptsize}
\begin{center}
\caption{Phase-resolved absorbed and unabsorbed fluxes of \source\ as measured by \ixpe. \label{tab:pulseprof}}
\begingroup
\renewcommand{\arraystretch}{1.4}
\setlength{\tabcolsep}{45pt}
\begin{tabular}{l l l l}
    \ & $2.0$--$7.5\,\mathrm{keV}$ & $2.0$--$4.0\,\mathrm{keV}$\ \ \ \ \ \ \  & $4.0$--$7.5\,\mathrm{keV}$ \\
\end{tabular}
\setlength{\tabcolsep}{14.5pt}
\begin{tabular}{c | c c | c c | c c }
\hline\hline
phase bin & $F_\mathrm{abs}$ & $F_\mathrm{unabs}$ & $F_\mathrm{abs}$ & $F_\mathrm{unabs}$ & $F_\mathrm{abs}$ & $F_\mathrm{unabs}$ \\
\hline
$0.00$--$0.14$ & $3.056^{+0.034}_{-0.035}$ & $3.112^{+0.035}_{-0.035}$ & $1.763^{+0.020}_{-0.020}$ & $1.812^{+0.020}_{-0.021}$ & $1.293^{+0.014}_{-0.015}$ & $1.300^{+0.015}_{-0.015}$ \\
$0.14$--$0.29$ & $4.231^{+0.038}_{-0.039}$ & $4.311^{+0.039}_{-0.040}$ & $2.562^{+0.023}_{-0.024}$ & $2.635^{+0.024}_{-0.024}$ & $1.669^{+0.015}_{-0.015}$ & $1.677^{+0.015}_{-0.015}$ \\
$0.29$--$0.43$ & $5.196^{+0.043}_{-0.044}$ & $5.295^{+0.044}_{-0.044}$ & $3.218^{+0.027}_{-0.027}$ & $3.308^{+0.027}_{-0.028}$ & $1.978^{+0.016}_{-0.017}$ & $1.988^{+0.016}_{-0.017}$ \\
$0.43$--$0.57$ & $5.102^{+0.042}_{-0.043}$ & $5.201^{+0.043}_{-0.044}$ & $3.141^{+0.026}_{-0.026}$ & $3.231^{+0.027}_{-0.027}$ & $1.960^{+0.016}_{-0.017}$ & $1.970^{+0.016}_{-0.017}$ \\
$0.57$--$0.71$ & $3.341^{+0.035}_{-0.035}$ & $3.405^{+0.036}_{-0.036}$ & $2.056^{+0.021}_{-0.022}$ & $2.113^{+0.022}_{-0.022}$ & $1.285^{+0.013}_{-0.014}$ & $1.292^{+0.013}_{-0.014}$ \\
$0.71$--$0.86$ & $5.132^{+0.044}_{-0.045}$ & $5.223^{+0.045}_{-0.046}$ & $2.906^{+0.025}_{-0.026}$ & $2.986^{+0.026}_{-0.026}$  & $2.226^{+0.019}_{-0.020}$ & $2.237^{+0.019}_{-0.020}$ \\
$0.86$--$1.00$ & $4.911^{+0.044}_{-0.045}$ & $4.997^{+0.045}_{-0.046}$ & $2.743^{+0.025}_{-0.025}$ & $2.817^{+0.025}_{-0.026}$ & $2.169^{+0.019}_{-0.020}$ & $2.178^{+0.020}_{-0.020}$ \\
\hline\hline
\end{tabular}
\endgroup
\end{center}
    \tablecomments{All values are in units of $10^{-11}\,\mathrm{erg}\,\mathrm{cm}^{-2}\,\mathrm{s}^{-1}$.}
\end{table*}

\section{Polarimetric analysis} \label{sec:polla}
We performed the polarimetric analysis of the \ixpe\ data inside {\sc xspec}, extracting the fluxes of the Stokes parameters $Q$ and $U$ from the processed photon lists following the same procedure discussed in the previous sections for the photon flux $I$. Then, following \citet{2017ApJ...838...72S}, we fit simultaneously the $I$, $Q$ and $U$ spectra of the three \ixpe\ DUs, convolving the spectral model, {\tt constant$\times$tbabs$\times$(bbodyrad$+$bbodyrad)}, with the {\tt polconst} model, and freezing the spectral parameters to the best fit values presented in section \ref{sec:spectral}. We checked that the results are fully consistent within the errors with those returned by the {\tt pcube} algorithm of the {\sc ixpeobssim} suite, uploading the most recently-updated response functions (v20260701). For this reason, 
in the following we will only discuss the spectro-polarimetric, weighted-analysis results obtained within {\sc xspec}.

\begin{figure}[!t]
\includegraphics[width=0.47\textwidth]{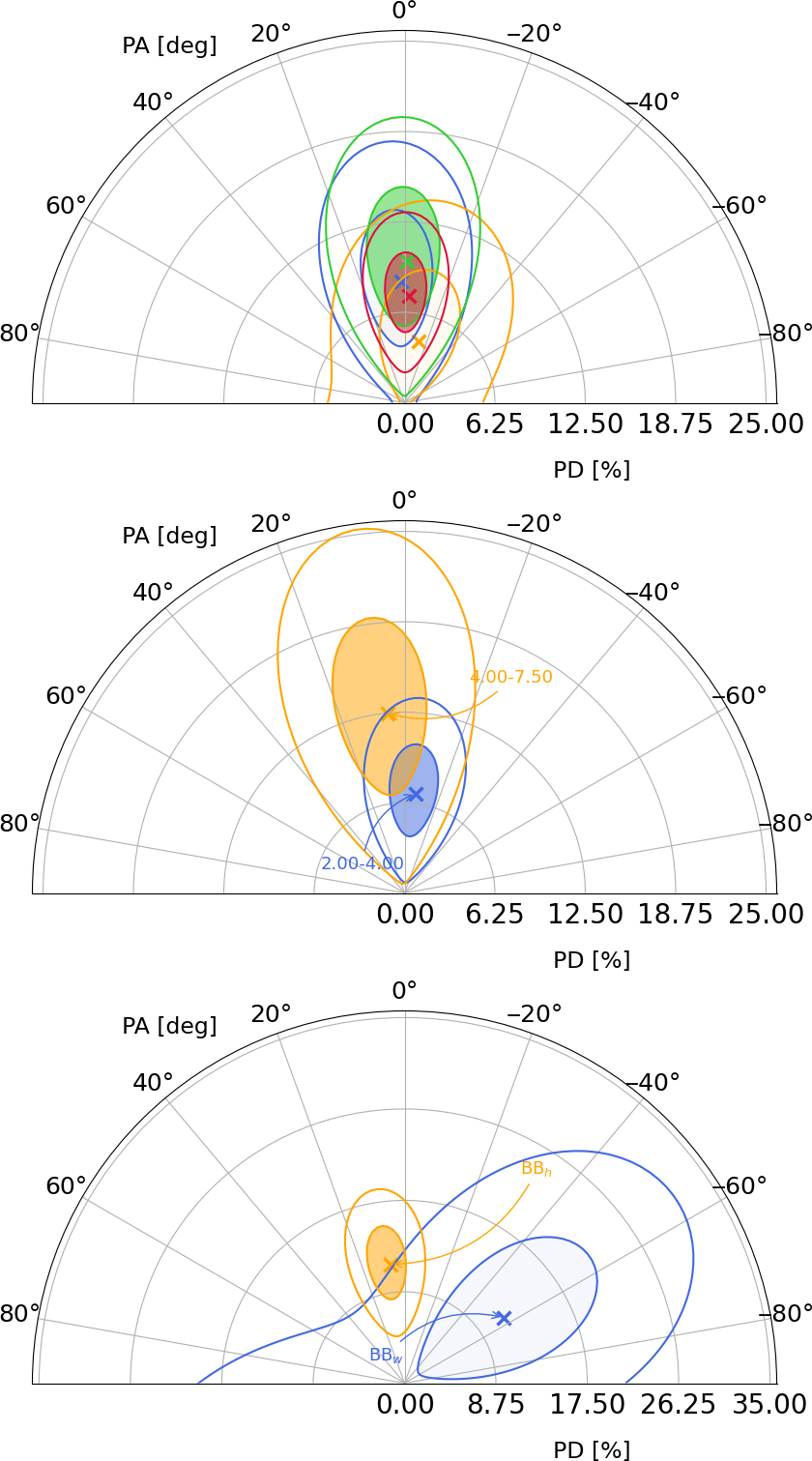}
\caption{Top panel: phase-integrated polarization degree and angle in the $2.0$--$7.5\,\mathrm{keV}$ detected by DU 1 (blue), 2 (orange) and 3 (green), together with the sum of the three DUs (red). Middle panel: phase-integrated polarization degree and angle in the $2.0$--$4.0\,\mathrm{keV}$ (blue) and in the $4.0$--$7.5\,\mathrm{keV}$ (orange) energy intervals. Bottom panel: phase-integrated polarization degree and angle in the $2.0$--$7.5\,\mathrm{keV}$ band relative to the warm BB (cyan) and to the hot BB (yellow) components. The filled and empty contours mark the $68\%$ and $99\%$ confidence intervals, respectively. 
\label{fig:pollapa}}
\end{figure}
We measured a significant phase-averaged degree of polarization ($4.19\sigma$ confidence level) in the $2.0$--$7.5\,\mathrm{keV}$ band for the total signal, summed over the three DUs, with $\mathrm{PD}=7.42\pm1.77\%$ \cite[above $\mathrm{MDP}_{99}=5.37\%$; see][]{2010SPIE.7732E..0EW} and polarization angle $\mathrm{PA}=-2.09^{+6.94}_{-6.93}$ deg (see Figure \ref{fig:pollapa}, top panel)\footnote{$\mathrm{PA}$ is positive East of the celestial North.}. Polarization is below  $\mathrm{MDP}_{99}$ for DU 1 and 2, and above for DU 3.
We also found significant polarization in the $2.0$--$4.0$ and $4.0$--$7.5\,\mathrm{keV}$ energy bands. The values of degree and angle of polarization measured in the two bands are consistent within $1\sigma$, anyway suggesting a (modest) change with energy (see Figure \ref{fig:pollapa}, middle panel, and Table \ref{tab:pollapa} for a summary of the values). To better investigate this variation of polarization with energy, we repeated the energy-dependent polarimetric fit across the entire $2.0$--$7.5\,\mathrm{keV}$ range, convolving the spectral components with a linearly energy-dependent polarization model ({\tt pollin} in {\sc xspec}), and alternately leaving the slopes of $\mathrm{PD}$ and $\mathrm{PA}$ either as free parameters or fixed at $0$ (thus reproducing a constant energy dependence). We found that additional degrees of freedom improve the $\chi^2/\mathrm{dof}$ ratio, which decreases from $492.63/496$ for constant $\mathrm{PD}$ and $\mathrm{PA}$, to $489.46/494$ when both $\mathrm{PD}$ and $\mathrm{PA}$ linearly vary with energy, and to $490.00/495$ for linearly dependent $\mathrm{PD}$ and constant $\mathrm{PA}$\footnote{The $\chi^2/\mathrm{dof}$ ratio increases to $491.92/495$ when constant $\mathrm{PD}$ and linearly dependent $\mathrm{PA}$ are considered.}. However, an F-test shows no statistically significant improvement relative to the constant scenario: the null-hypothesis probability drops at $\approx10\%$ only when $\mathrm{PA}$ is the sole constant parameter ($\mathrm{nhp}=10.37\%$), whereas $\mathrm{nhp}$ ranges from $\approx20\%$ to $\approx40\%$ in the other cases. Nevertheless, the marginal improvement obtained for linearly dependent $\mathrm{PD}$ and constant $\mathrm{PA}$ is suggestive of a possible weak increase of the polarization degree with energy.

\begin{table}[!t]
\tabletypesize{\scriptsize}
\begin{center}
\caption{Phase-averaged polarization measured by \ixpe\ in different energy bands. \label{tab:pollapa}}
\begingroup
\setlength{\tabcolsep}{8pt}
\renewcommand{\arraystretch}{1.6}
\begin{tabular}{c c c c}
\hline\hline
\  & $2.0$--$4.0\,\mathrm{keV}$ & $4.0$--$7.5\,\mathrm{keV}$ & $2.0$--$7.5\,\mathrm{keV}$ \\
\hline
$\mathrm{PD}$ [$\%$] & $6.92^{+2.05}_{-2.05}$ & $12.47^{+3.92}_{-3.92}$ & $7.42^{+1.77}_{-1.77}$ \\
$\mathrm{MDP}_{99}$ [$\%$] & $6.22$ & $11.90$ & $5.38$ \\
$\mathrm{PA}$ [$\mathrm{deg}$] & $-6.33^{+8.68}_{-8.62}$ & $5.50^{+9.17}_{-9.09}$ & $-2.09^{+6.94}_{-6.93}$ \\
$\chi^2/\mathrm{dof}$ & $186.59/199$ & $292.71/286$ & $492.63/496$ \\
\hline\hline
\end{tabular}
    \endgroup
    \end{center}
\end{table}
\begin{table}[!t]
\tabletypesize{\scriptsize}
\begin{center}
\caption{Phase- and energy-integrated polarization degree and angle associated with the two spectral components discussed in section \ref{sec:spectral} (see bottom panel of Figure \ref{fig:pollapa}). \label{tab:pollaspecpol}}
\begingroup
\setlength{\tabcolsep}{17pt}
\renewcommand{\arraystretch}{1.6}
\begin{tabular}{c c c}
\hline\hline
\  & Warm BB & Hot BB \\
\hline
$\mathrm{PD}$ [$\%$] & $11.35^{+6.42}_{-6.42}$ & $11.44^{+2.30}_{-2.30}$ \\
$\mathrm{MDP}_{99}$ [$\%$] & $19.49$ & $6.98$ \\
$\mathrm{PA}$ [$\mathrm{deg}$] & $-56.34^{+17.24}_{-17.24}$ & $6.62^{+5.82}_{-5.82}$ \\
\hline\hline
\end{tabular}
    \endgroup
    \end{center}
\end{table}
Keeping the values of the spectral parameters fixed to those reported in Table \ref{tab:specpa}, we performed a spectro-polarimetric analysis by convolving each single BB component with the {\tt polconst} model ($\chi^2/\mathrm{dof}=490.23/496$). The results are reported in Table \ref{tab:pollaspecpol} and shown in the bottom panel of Figure \ref{fig:pollapa}. The polarization degree associated with the hot blackbody component was found to be $11.44\pm2.30\%$ (above $\mathrm{MDP}_{99}=6.98\%$), with a polarization angle $6^\circ.62\pm5^\circ.82$, both consistent within $1\sigma$ with the phase-averaged polarization measured at energies $4.0$--$7.5\,\mathrm{keV}$. On the other hand, the polarization estimate for the warm thermal component is $\mathrm{PD}=11.35\pm6.42\%$, below $\mathrm{MDP}_{99}=19.49\%$, with $\mathrm{PA}=-56^\circ.34\pm17^\circ.24$. This differs by $\approx2.5\sigma$ from 
the value ($\approx-6^\circ$) detected between $2.0$ and $4.0\,\mathrm{keV}$ .

\begin{figure}[!t]
\includegraphics[width=0.47\textwidth]{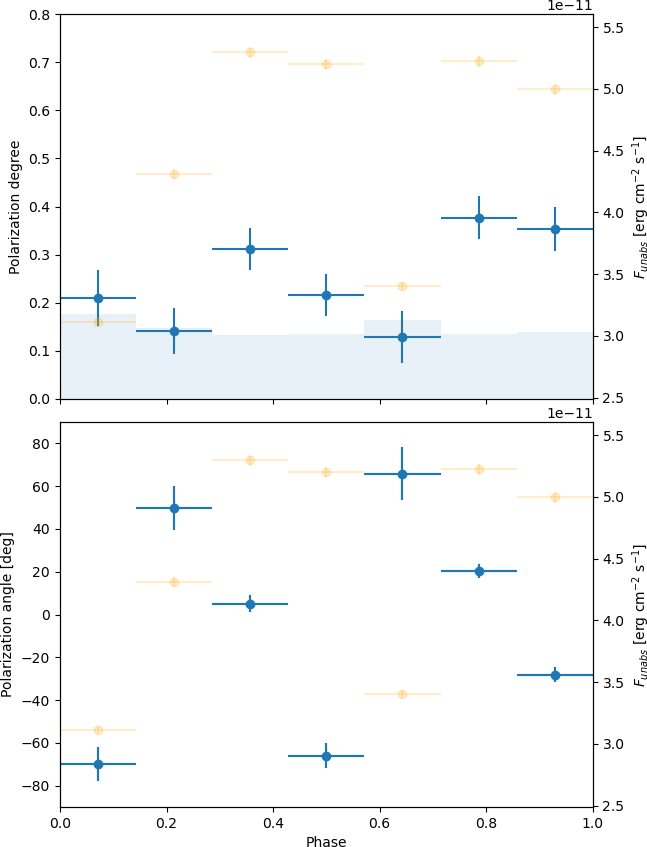}
\caption{Phase-dependent polarization degree (top) and angle (bottom) in the $2.0$--$7.5\,\mathrm{keV}$ band (blue dots with $1\sigma$ error bars). The corresponding $\mathrm{MDP}_{99}$ value in each phase bin (shaded histogram) is reported in the top panel. 
In both panels, the $2.0$--$7.5\,\mathrm{keV}$ light curve (see Figure \ref{fig:lightcurves}) is also shown for comparison (yellow points with error bars).
\label{fig:pollapr}}
\end{figure}

In order to explore how the polarization properties of \source\ change with the spin phase, we divided the rotational period into seven, equally-spaced phase intervals, as discussed in section \ref{sec:spectral} for the phase-dependent spectral analysis. This choice is a good compromise between tracing the phase variation of the degree and angle of polarization, and having detections above $\mathrm{MDP}_{99}$ in as many phase bins as possible. Figure \ref{fig:pollapr} shows the phase-dependent behavior of $\mathrm{PD}$ and $\mathrm{PA}$ obtained by integrating the Stokes parameters in the $2.0$--$7.5\,\mathrm{keV}$ band (numerical values are reported in Table \ref{tab:pollapr}). 

\begin{table*}[ht!]
\tabletypesize{\scriptsize}
\begin{center}
\caption{Phase-resolved polarization degree and angle as measured by \ixpe\ in three different energy bands$^\mathrm{(a)}$. \label{tab:pollapr}}
\begingroup
\renewcommand{\arraystretch}{1.4}
\begin{tabular}{c c c}
$2.0$--$7.5\,\mathrm{keV}$\ \ \ \ \ \ \ \ \ \ \ \ \ \ \ \ \ \ \ \ \ \ \ \ \ \ &\ \ \ \ \ \ \ \ \ \ $2.0$--$4.0\,\mathrm{keV}$\ \ \ \ \ \ \ \ \ \ \ \ \ \ \ \ \ \ \  &\ \ \ \ \ \ \ \ \ \ \ \ \ \ \ $4.0$--$7.5\,\mathrm{keV}$ \\
\end{tabular}
\setlength{\tabcolsep}{3pt}
\begin{tabular}{c | c c c | c c c | c c c }
\hline\hline
phase bin & $\mathrm{PD}$ [$\%$] & $\mathrm{MDP}_{99}$ [$\%$] & $\mathrm{PA}$ [$\mathrm{deg}$] & $\mathrm{PD}$ [$\%$] & $\mathrm{MDP}_{99}$ [$\%$] & $\mathrm{PA}$ [$\mathrm{deg}$] & $\mathrm{PD}$ [$\%$] & $\mathrm{MDP}_{99}$ [$\%$] & $\mathrm{PA}$ [$\mathrm{deg}$] \\
\hline
$0.00$--$0.14$ & $21.05^{+5.83}_{-5.83}$ & $17.69$ & $-69.93^{+8.05}_{-8.03}$ & $9.08^{+6.74}_{-6.74}$ & $20.46$ & $-61.51^{+23.79}_{-23.92}$ & $57.54^{+12.99}_{-12.99}$ & $39.42$ & $-78.06^{+6.42}_{-6.34}$ \\
$0.14$--$0.29$ & $14.10^{+4.84}_{-4.84}$ & $14.70$ & $49.77^{+10.14}_{-10.12}$ & $4.49^{+5.58}_{-4.49}$ & $16.92$ & $51.02^{+38.98}_{-51.02}$ & $47.62^{+10.80}_{-10.80}$ & $32.78$ & $46.90^{+6.65}_{-6.65}$ \\
$0.29$--$0.43$ & $31.20^{+4.37}_{-4.37}$ & $13.26$ & $5.08^{+4.02}_{-4.02}$ & $29.97^{+5.05}_{-5.05}$ & $15.34$ & $2.98^{+4.78}_{-4.77}$ & $38.25^{+9.65}_{-9.64}$ & $29.30$ & $14.68^{+7.43}_{-7.60}$ \\
$0.43$--$0.57$ & $21.64^{+4.43}_{-4.43}$ & $13.45$ & $-65.94^{+5.90}_{-5.91}$ & $21.48^{+5.08}_{-5.08}$ & $15.43$ & $-65.60^{+6.81}_{-6.85}$ & $29.39^{+10.04}_{-10.04}$ & $30.48$ & $-56.52^{+10.26}_{-10.05}$ \\
$0.57$--$0.71$ & $12.84^{+5.44}_{-5.44}$ & $16.52$ & $65.70^{+12.58}_{-12.40}$ & $12.45^{+6.24}_{-6.24}$ & $18.93$ & $70.31^{+15.31}_{-14.93}$ & $13.25^{+12.23}_{-12.23}$ & $37.12$ & $44.62^{+33.23}_{-25.80}$ \\
$0.71$--$0.86$ & $37.66^{+4.43}_{-4.43}$ & $13.46$ & $20.38^{+3.38}_{-3.38}$ & $30.38^{+5.21}_{-5.21}$ & $15.80$ & $20.85^{+4.93}_{-4.94}$ & $61.37^{+9.39}_{-9.38}$ & $28.49$ & $17.82^{+4.37}_{-4.35}$ \\
$0.86$--$1.00$ & $35.33^{+4.56}_{-4.56}$ & $13.83$ & $-28.06^{+3.70}_{-3.70}$ & $26.03^{+5.37}_{-5.37}$ & $16.31$ & $-27.64^{+5.95}_{-5.93}$ & $62.91^{+9.43}_{-9.42}$ & $28.61$ & $-26.57^{+4.33}_{-4.39}$ \\
\hline\hline
\end{tabular}
\endgroup
\end{center}
    \tablecomments{\\
    (a) -- Errors are at $1\sigma$ confidence level (the corresponding values of $\mathrm{MDP}_{99}$ are also reported).
    }
\end{table*}
The polarization degree is found to be below $\mathrm{MDP}_{99}$ in only two out of seven phase bins, corresponding to the rising phase towards the first peak (at phase $0.14$--$0.29$) and to the minimum between the two peaks (at phase $0.57$--$0.71$). However, even considering the upper limits at $3\sigma$ for $\mathrm{PD}$ in these bins, the phase-dependent behavior exhibits a clear double-peaked profile ($p\approx2.14\times10^{-4}$ for the fit with a constant function), similar to that of the light curve, with values at the two peaks of $\approx30\%$ and $\approx40\%$, respectively. By contrast, the polarization angle appears to be completely uncorrelated with the pulse profile, showing a continuous swing that covers the entire range of variation (from $-90^\circ$ to $+90^\circ$). We note that this behavior is typical of the rotating vector model \cite[RVM,][]{1969ApL.....3..225R,1970Natur.225..612K} for particular combinations of geometrical angles (see the discussion below).

\begin{figure}[!t]
\includegraphics[width=0.47\textwidth]{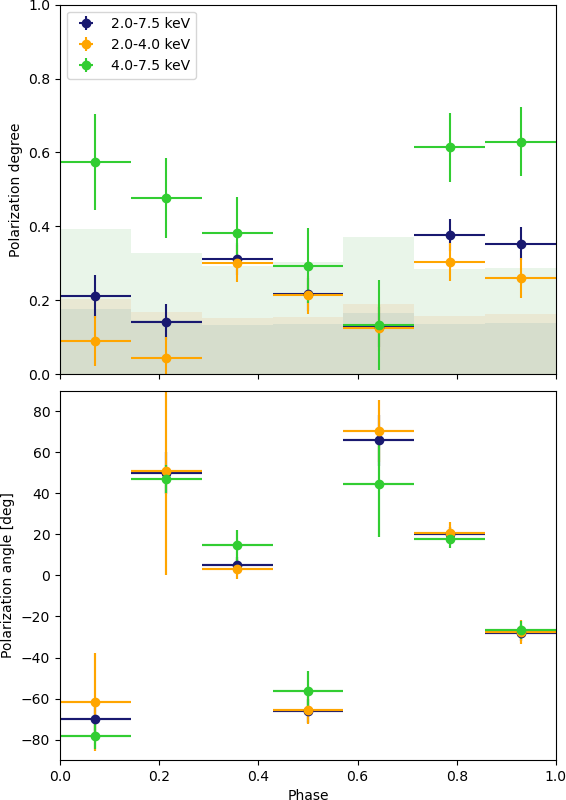}
\caption{Phase-dependent polarization degree (top) and angle (bottom) in the $2.0$--$7.5$ (blue), $2.0$--$4.0$ (orange) and $4.0$--$7.5\,\mathrm{keV}$ (green) energy bands. The same color code is used to mark the corresponding values of the $\mathrm{MDP}_{99}$ in each phase bin and energy band. Error bars mark $1\sigma$ errors.
\label{fig:pollapr_multiband}}
\end{figure}

The phase-dependent behavior of $\mathrm{PD}$ and $\mathrm{PA}$ in the $2.0$--$4.0$ and $4.0$--$7.5\,\mathrm{keV}$ bands shows that the polarization increases with increasing energy, in agreement with the phase-integrated results discussed above (see Table \ref{tab:pollapr} and Figure \ref{fig:pollapr_multiband}). The polarization degree at $2.0$--$4.0\,\mathrm{keV}$ appears to be consistent with that in the entire $2.0$--$7.5\,\mathrm{keV}$ band, with the values of $\mathrm{PD}$ in the two peaks consistent within $1\sigma$. On the other hand, at $4.0$--$7.5\,\mathrm{keV}$ the polarization is significantly higher (the best-fit value obtained with a constant function is $45.70\pm3.77\%$, compared to $20.46\pm2.00\%$ in the lower energy band). At variance with the behavior at low energies, the polarization degree at high energies appears to attain a minimum in correspondence to the first peak (at $38.3\pm9.6\%$), while a significantly higher value is reached around the second peak ($62.9\pm9.4\%$). 
The marked increase of the phase-dependent polarization at high energies is not shared by the corresponding phase-averaged quantity (see Figure \ref{fig:pollapa}). This can be easily explained by noting the rapid phase variation of the polarization angle, which maintains (within the errors) the same trend observed in the $2.0$--$7.5\,\mathrm{keV}$ also in the $2.0$--$4.0$ and $4.0$--$7.5\,\mathrm{keV}$ bands (see the lower panel in Figure \ref{fig:pollapr_multiband}). This is a further confirmation that the polarization direction is uniquely determined by the source viewing geometry, as expected from the rotating vector model.

\begin{figure}[!t]
\includegraphics[width=0.47\textwidth]{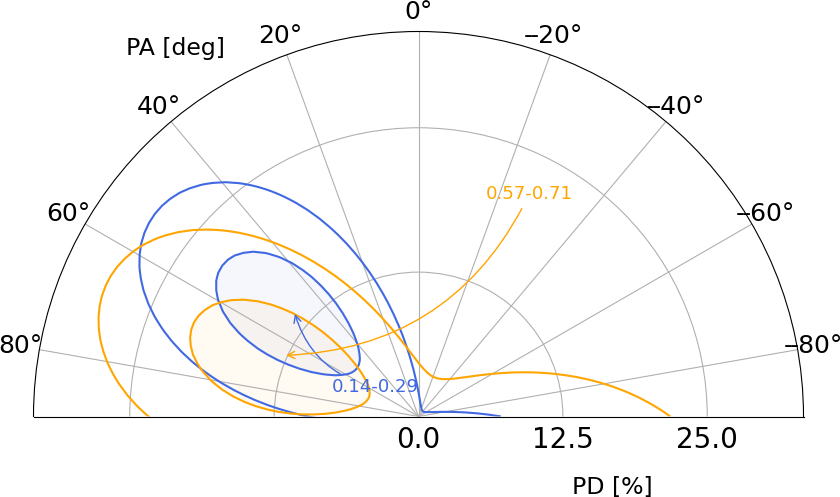}
\caption{Polarization degree and angle measured in the $2.0$--$7.5\,\mathrm{keV}$ band by \ixpe\ in the phase bins where $\mathrm{PD}$ results below $\mathrm{MDP}_{99}$ (see Table \ref{tab:pollapr}). Details as in Figure \ref{fig:pollapa}.
\label{fig:polarplotaperti}}
\end{figure}
To test this scenario quantitatively and  constrain the geometry of the system, we fitted the $2.0$--$7.5\,\mathrm{keV}$, phase-dependent polarization angle with the RVM model \cite[][]{1970Natur.225..612K},
\begin{equation} \label{eqn:rvm}
    \tan(\mathrm{PA} -\Psi) =\frac{\sin\xi\sin[\pm(\gamma+\gamma_0)]}{\sin\chi\cos\xi-\cos\chi\sin\xi\cos(\gamma+\gamma_0)}\,,
\end{equation}
where $\chi$ and $\xi$ are the angles between the observer's line-of-sight (LOS) and star magnetic-dipole axis with respect to the star spin axis, respectively, $\gamma$ is the rotational phase, $\gamma_0$ is an initial phase and $\Psi$ is an angular offset. Although the polarization falls below  $\mathrm{MDP}_{99}$ in two phase bins, the polarization angle is still well characterized at the $1\sigma$ confidence level (only the $99\%$ contours include zero polarization; see Figure \ref{fig:polarplotaperti}). We then decided to also use the values of $\mathrm{PA}$ in these two bins with $1\sigma$ confidence intervals in our fit with the RVM model. The fit is satisfactory ($\chi^2/\mathrm{dof}=1.698/3$) and returns $\chi=28^\circ.20\pm9^\circ.62$, $\xi=108^\circ.94\pm9^\circ.21$, $\Psi=-21^\circ.72\pm12^\circ.40$ and $\gamma_0=32^\circ.30\pm11^\circ.83$. We checked \emph{a posteriori} that the best-fit parameters obtained assuming that the polarization angle is not constrained in the bins below $\mathrm{MDP}_{99}$ are fully consistent, within $1\sigma$, with those derived with the approach described above. 

\begin{figure}[!t]
\includegraphics[width=0.47\textwidth]{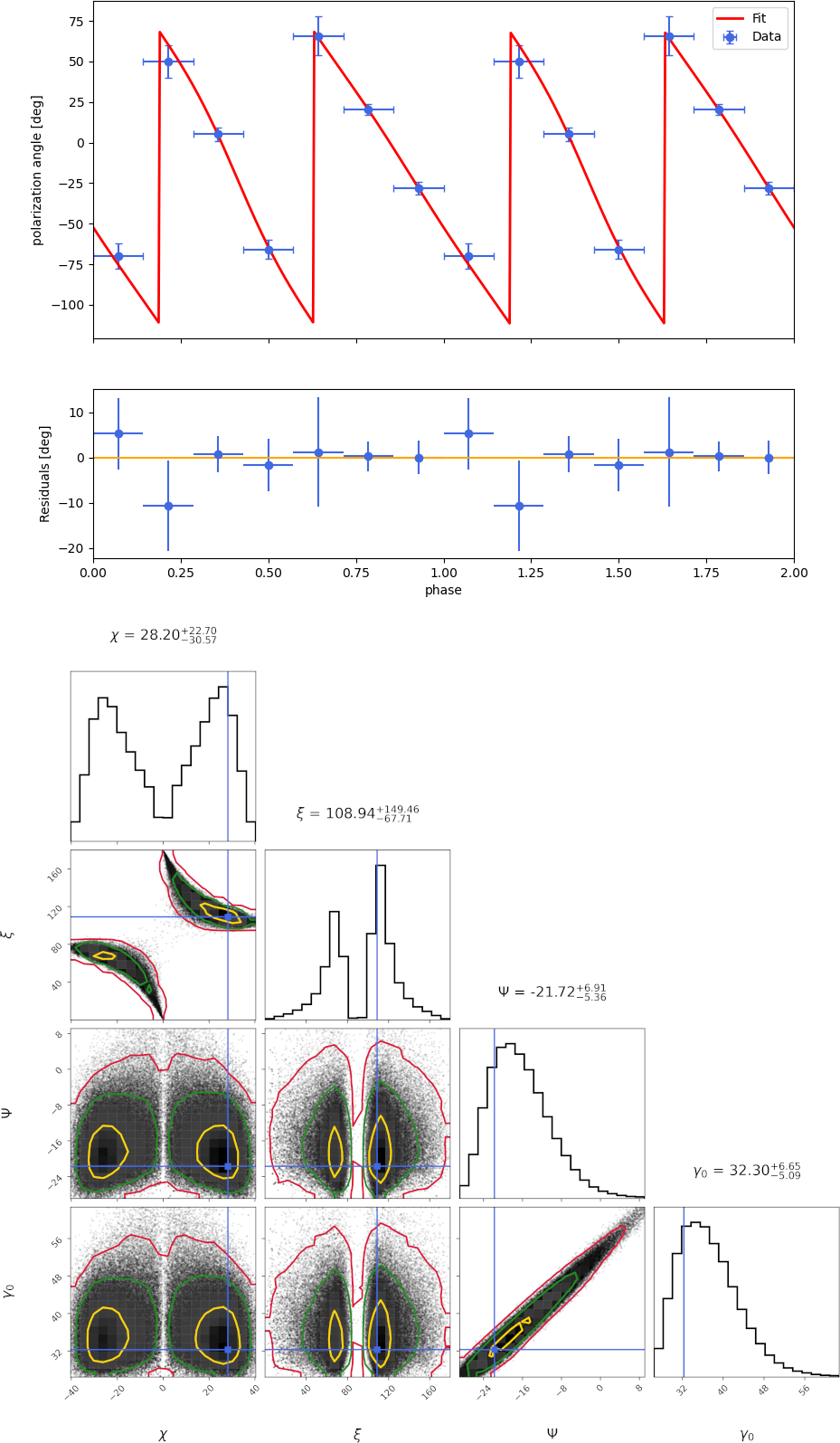}
\caption{Top: \ixpe\ $2.0$--$7.5\,\mathrm{keV}$ polarization angle (blue points with error bars) fitted with the RVM model (see equation \ref{eqn:rvm}) (red solid line). The errors at $1\sigma$ have been reported also for the phase bins where $\mathrm{PD}$ falls below $\mathrm{MDP}_{99}$ (see text for more details). Bottom: marginalized posterior distribution in the region around the best-fitting solution. 
The yellow, green and red contours show the $1$, $2$ and $3\sigma$ confidence regions, respectively, while the blue solid lines mark the best-fit values obtained by minimizing the $\chi^2$. The (joint) maximum \emph{a posteriori} estimates are reported above each likelihood histogram. 
\label{fig:rvmfit}}
\end{figure}
To fully characterize the posterior probability distribution of the RVM geometrical parameters, we performed an extensive MCMC-based search of the $\chi$--$\xi$ plane (using flat priors extended over the entire range of variation of the different parameters), probing all regions of maximum likelihood. The search returned 16 formally distinct best-fit solutions, all characterized by identical goodness of fit ($\chi^2/\mathrm{dof}=1.698/3$). Actually, these solutions are not independent since they are related by the transformations $\chi,\xi\rightarrow\chi,\xi\pm180^\circ$, together with the trivial periodicity of $360^\circ$ in both angles. These transformations leave the predicted RVM curve unchanged, indicating that all 16 solutions correspond to a single underlying geometry. We therefore adopt $\chi=28^\circ.20$ and $\xi=108^\circ.94$, with $\Psi=-21^\circ.72$ and $\gamma_0=32^\circ.30$, as our reference solution. To explore the local posterior probability density, we ran a targeted MCMC chain, using 30,000 walkers initialized around this reference solution (see Figure \ref{fig:rvmfit}). The residual degeneracy reflects the well-known properties of the rotating vector model, i.e. the intrinsic $180^\circ$ ambiguity in the definition of the polarization position angle, the inability of polarimetric data alone to distinguish between the two magnetic poles, as well as the north/south ambiguity of the viewing geometry and of the direction of rotation of the star projected on the plane of the sky. 

As a further test, we also fitted the phase-resolved polarization angle in the $4.0$--$7.5\,\mathrm{keV}$ band, adopting the same procedure described above for the entire energy range. Although the fit yields $\chi^2/\mathrm{dof}=6.724/3$, with the third phase bin providing the dominant contribution to the total $\chi^2$ given the limited number of independent phase bins, the best-fit geometry ($\chi=21^\circ.82\pm16^\circ.93$ and $\xi=106^\circ.45\pm12^\circ.52$, with $\Psi=-22^\circ.75\pm26^\circ.02$ and $\gamma_0=31^\circ.44\pm26^\circ.64$) is nevertheless fully consistent with that derived in the $2.0$--$7.5\,\mathrm{keV}$ band. 

\section{Discussion}
Outbursts, events characterized by an enhancement of the persistent flux up to three orders of magnitude with respect to quiescence, have been the primary channel for discovering new magnetars, which are otherwise often too faint for detection with all-sky monitors. This was the case of \source, the first magnetar observed in a full-fledged outburst since the launch of \ixpe\ in December 2021. Triggered by the initial {\it EP} detection, a prompt \ixpe\ ToO observation was carried out, allowing for the first X-ray polarimetric study of a  magnetar outburst. 

Although \ixpe\ targeted \source\ for a (relatively) short exposure time ($\approx 200\,\mathrm{ks}$), the observed flux in the $2$--$8\,\mathrm{keV}$ energy range ($\approx 5\times10^{-11}\,\mathrm{erg\,cm^{-2}\,s^{-1}}$) was high enough to provide a robust polarization measurement, with the signal remaining essentially unaffected by background contamination nearly throughout the entire \ixpe\ band ($2.0$--$7.5\,\mathrm{keV}$). Our phase-integrated spectral analysis confirms the presence of at least two thermal components with different temperatures and radii, as first reported by \citet[see Table \ref{tab:specpa}]{2026ATel17870....1R}. As shown in Figure \ref{fig:spectrumphaseint} (left panel), the warm BB component contributes to the total spectrum up to $\approx4\,\mathrm{keV}$ where the hot BB takes over. 
We therefore investigated the spectral and polarization properties in the $2.0$--$4.0\,\mathrm{keV}$ and $4.0$--$7.5\,\mathrm{keV}$ bands separately. 

The temperatures and emission radii of both thermal components as derived from the phase-resolved spectral analysis remain consistent with those obtained from the phase-averaged spectrum 
and favor a scenario in which radiation originates from localized regions of the stellar surface, providing a natural geometric explanation for the moderately high pulsed fractions observed ($\approx25\%$, with a peak of $\approx30\%$ at $2.0$--$4.0\,\mathrm{keV}$). 
This is further confirmed by the persistence of the double-peak structure in the pulse profile at all energies (see Figure \ref{fig:lightcurves}), with only minimal changes in the relative amplitude of the peaks. Such a localized emission geometry prevents treating high polarization degrees as unambiguous signatures of quantum electrodynamic (QED) effects like vacuum birefringence. 

The modest energy- and phase-averaged polarization degree ($\approx 7\%$) seems to be essentially driven by low-energy photons ($2.0$--$4.0\,\mathrm{keV}$). The higher polarization ($\approx13\%$) measured at $4.0$--$7.5\,\mathrm{keV}$, although still consistent with the low-energy value within $1\sigma$, suggests a possible increase with energy, while maintaining the same polarization direction in the sky (see section \ref{sec:polla}). This increasing trend is further supported by the phase-resolved analysis, which reveals a higher average polarization of $\approx50\%$ at high energies, compared to $\approx 20\%$ at lower energies.

The phase-dependent behavior of the polarization angle is well fit by the RVM, while the polarization degree shows a clear phase dependence associated with the double-peaked pulse profile (although the phase-dependent shape of $\mathrm{PD}$ changes with energy), as found in other persistent magnetars observed by \ixpe\ \cite[][]{2022Sci...378..646T,2023ApJ...944L..27Z,2024MNRAS.52712219H,2026ApJ..1002..102T}. This finds a natural explanation in the presence of  vacuum birefringence effects in the star magnetosphere, where the observed polarization degree reflects the polarization properties of radiation at the emission, whereas the polarization direction is frozen at the polarization-limiting radius, which is expected to be much larger than the stellar radius for magnetars \cite[][]{2003MNRAS.342..134H,2015MNRAS.454.3254T}. No $90^\circ$ shift in the polarization angle across the \ixpe\ band is visible (see Table \ref{tab:pollapa}), suggesting that the same polarization mode (either ordinary, O, or extraordinary, X) dominates at all energies\footnote{In a highly magnetized environment, radiation is linearly polarized either parallel (ordinary mode) or perpendicular (extraordinary mode) to the plane defined by the photon wave vector and the local magnetic field \citep[see also \citealt{1997JPhA...30.6485H} and \citealt{2006RPPh...69.2631H} for a comprehensive review]{1978JETPL..27..305G}.}. For the geometry favored by the RVM model (inclination of the line of sight and of the magnetic dipole axis relative to the star spin axis of $\chi\approx25^\circ$ and $\xi\approx110^\circ$, respectively) the polarization angle continuously swings across the entire $-90^\circ,+90^\circ$ range during a cycle. In this case, averaging the Stokes parameters over the rotational phase results in a low polarization degree, even accounting for vacuum birefringence \cite[see][Figure 7, in particular]{2015MNRAS.454.3254T}. This provides a self-consistent explanation for the low phase-averaged polarization degree at all energies, in contrast to the much higher phase-resolved values, and shows that, under these conditions, the intrinsic polarization degree is better tracked by phase-resolved polarimetry.

\begin{figure}[!t]
\includegraphics[width=0.47\textwidth]{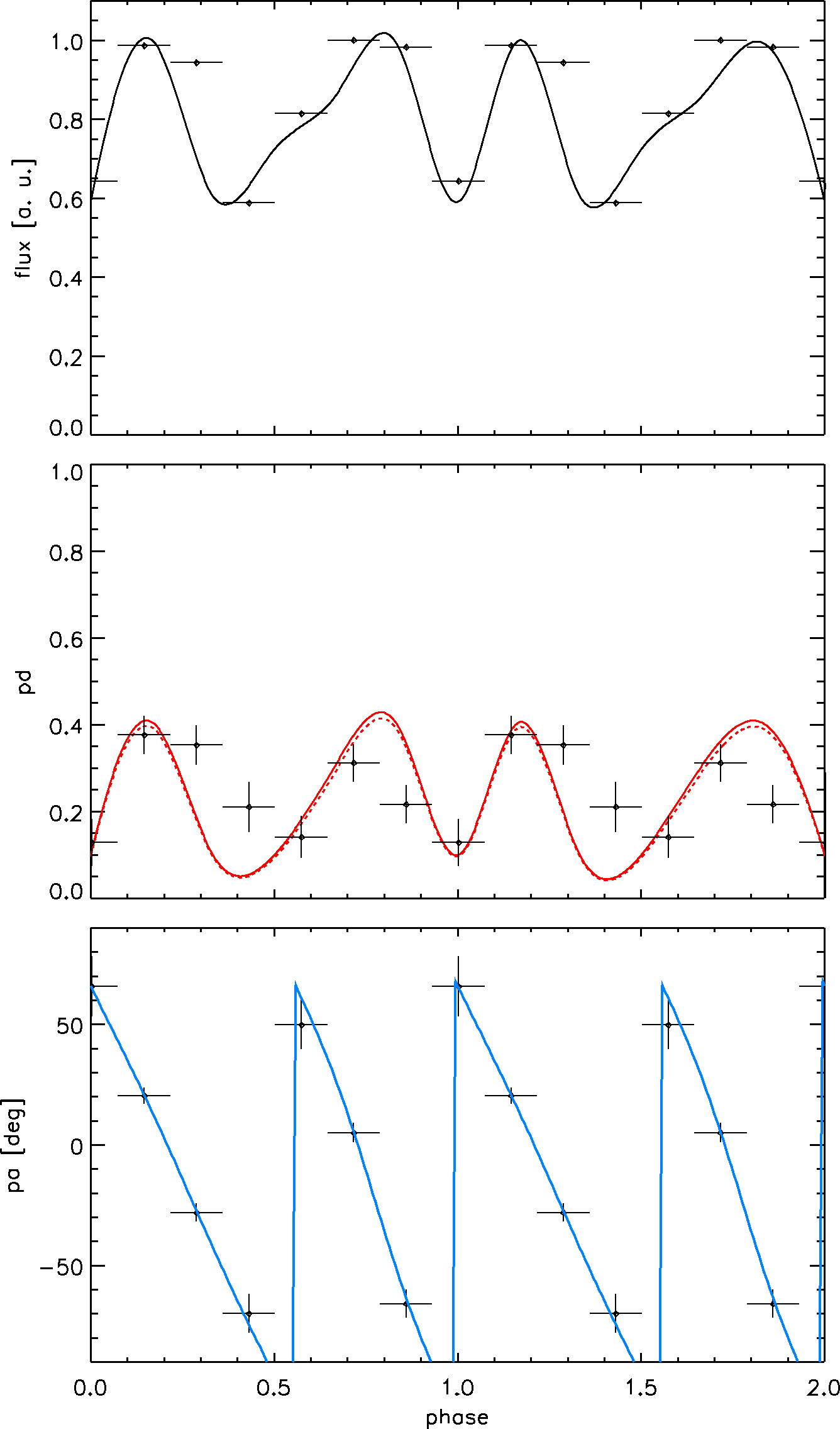}
\caption{Phase-dependent $2.0$--$7.5\,\mathrm{keV}$ (normalized) flux (top, black lines), polarization degree (center, red lines) and polarization angle (bottom, blue lines) simulated with the ray-tracing code, including (solid lines) or neglecting (dashed lines) vacuum birefringence effects. The black diamonds with $1\sigma$ error bars show the \ixpe\ data. 
\label{fig:RTsimulation}}
\end{figure}
The high temperature (exceeding $1\,\mathrm{keV}$) and high polarization ($\gtrsim50\%$, with peaks of $\approx65\%$) of the hot thermal component point to the presence of a magnetized gaseous atmosphere, which, for the magnetic field strength inferred for \source\ ($\approx4\times10^{14}\,\mathrm{G}$ at the poles), is expected to emit radiation predominantly polarized in the X-mode \cite[][]{2020MNRAS.492.5057T,2024MNRAS.528.3927K}. Given the small emission radius inferred from the spectral analysis, the region affected by  atmospheric reprocessing is likely restricted to a small hot spot, possibly heated during the outburst \citep[see][for magneto-thermal simulations of magnetar outbursts]{2022ApJ...936...99D,2025A&A...701A.229D}. The low polarization ($\approx20\%$) detected at lower energies and associated with a distinct spectral component suggests the presence of a different emission mechanism. The temperature of the warm blackbody ($\approx0.5\,\mathrm{keV}$) and the value of the spin-down magnetic field are compatible with the presence of a condensed region on (part) of the star surface \cite[][see also \citealt{2006PhRvA..74f2508M,2007MNRAS.382.1833M,2012A&A...546A.121P}]{2020MNRAS.492.5057T}. The expected degree of polarization is substantially lower for emission from a magnetic condensate ($\lesssim20$--$30\%$), and the dominant polarization mode can be ordinary or extraordinary, depending on the viewing geometry \cite[][]{2024Galax..12....6T}. The close alignment of the phase-averaged polarization directions in the $2.0$--$4.0\,\mathrm{keV}$ and $4.0$--$7.5\,\mathrm{keV}$ bands, alongside the absence of sharp transitions in the phase-dependent polarization angle, seemingly points toward emission in the same mode as at high energies (i.e. the extraordinary one). 

In order to qualitatively test this scenario without attempting a detailed fit to the data, we produced  numerical simulations using the ray-tracing code described in \citet{2015MNRAS.454.3254T}. The emission is assumed to originate from two antipodal circular caps with semi-apertures $\approx9^\circ$, and from an elongated spot centered at  magnetic colatitude $\theta_\mathrm{h}=27^\circ$ 
with angular semi-apertures of $\approx2^\circ$ and $\approx1^\circ$. These values are broadly consistent with the emission radii inferred from the phase-resolved spectral analysis. The star mass and radius are taken to be $1.4\,M_\odot$ and $13\,\mathrm{km}$, respectively, with the surface sampled with a $\Theta\times\Phi=150\times150$ angular mesh, and we assume a purely dipolar magnetic field \cite[corrected for general relativistic effects, see][]{1996ApJ...473.1067P}, with a polar strength of $4\times10^{14}\,\mathrm{G}$. Radiation is assumed to follow pure blackbody distributions, with a temperature of $0.49\,\mathrm{keV}$ and $10\%$ polarization in the X-mode for the antipodal regions and a temperature of $1.1\,\mathrm{keV}$, $80\%$ polarized in the X-mode for the spot. The rest of the stellar surface emits unpolarized radiation at a lower temperature ($0.12\,\mathrm{keV}$). General relativistic effects (ray bending and gravitational redshift) are fully accounted for, and all temperatures and energies are those measured at infinity. Finally, the viewing geometry is set to the values derived from the RVM fit, i.e. $\chi=28^\circ.21$ and $\xi=108^\circ.90$. 

The results of the numerical simulation, integrated throughout the $2.0$--$7.5\,\mathrm{keV}$ energy band, are reported in Figure \ref{fig:RTsimulation}, where the \ixpe\ data have been superimposed for visual comparison. Even without aiming to  reproduce quantitatively the observed behavior, the adopted emission geometry proves capable to reproduce the double-peaked profile of the light curve, together with a similarly double peaked polarization degree (in phase with the flux). Indeed, the high polarization coming from the isolated hot spot is primarily responsible for the phase-dependent behavior of the polarization degree, while the weakly polarized radiation coming from the larger antipodal regions reduces the overall polarization from the intrinsic $80\%$ to the observed $30$--$40\%$. In particular, a closer look at the middle panel of Figure \ref{fig:RTsimulation} shows that the difference between the models obtained with (solid lines) and without (dashed lines) accounting for vacuum birefringence effects is  marginal, the more so when compared with the $1\sigma$ error bars of the data\footnote{The QED-on and QED-off polarization angle models are indistinguishable in the bottom panel of Figure \ref{fig:RTsimulation}, owing to the simplifying choice of a purely dipolar field at the surface of the star.}. This confirms that, as mentioned above, vacuum birefringence effects cannot be unambiguously probed simply by assessing the polarization degree when radiation originates from such restricted regions. We note that this simulation was obtained assuming the warm and hot thermal components to be polarized in the same mode, and that the phase dependence of the polarization degree does not yet track the observed dataset in full detail. Achieving higher precision is left to future work, with the implementation of self-consistent atmospheric and condensed surface models for the isolated and antipodal regions, thus incorporating a non-trivial angular dependence for both photon distributions and polarization, which is missing at present.

The outburst of \source\ highlighted once again the potential of X-ray polarimetry to provide crucial information of magnetar sources, complementary to those obtained from spectral and timing analyses alone. Unfortunately, in the case at hand the limited exposure time prevented us from tracking the temporal evolution of the polarization properties as the outburst decayed. Such monitoring would be particularly valuable for investigating whether these events are accompanied by changing magnetic topology  \cite[see e.g.][]{2021MNRAS.502.1549T,2025SCPMA..6819505G}. In this respect, the development of next-generation X-ray polarimeters will be instrumental in enabling such investigations. Missions such as {\it eXTP} \cite[][with launch expected in 2030]{2025SCPMA..6819502Z}, with its substantial improvement in soft X-ray sensitivity, and {\it EXPO} \cite[][proposed for ESA's M8 mission]{2026arXiv260820898S}, which offer the possibility of measuring polarization at higher energies, will make it possible to extend this type of investigation without necessarily requiring prohibitively long exposure times.


\section*{Acknowledgments}
We thank the anonymous referee for their useful comments and suggestions, which helped to improve this work. This research was supported by the International Space Science Institute (ISSI) in Bern, through the International Team project 25-657 ``Polarimetric Insights into Extreme Magnetism''. The work of R. Ta., L. M., R. Tu. and R.M.E. K. is partially funded through the PRIN grant 2022 - project 2022LWPEXW - ``An X-ray view of compact objects in polarized light'', European Union funding - Next Generation EU, Mission 4 Component 1, CUP C53D23001180006.
F.C.Z. is supported by a Ram\'on y Cajal fellowship (grant agreement RYC2021-030888-I). F.C.Z., N.~R., and Y.L.W. acknowledge support from the Spanish grant PID2023-153099NA-I00.
A.B. acknowledges support through the European Space Agency (ESA) research fellowship programme.
This work uses data obtained with the Einstein Probe, a space mission supported by the Strategic Priority Program on Space Science of the Chinese Academy of Sciences, in collaboration with the European Space Agency, the Max-Planck-Institute for extraterrestrial Physics (Germany), and the Centre National d’Etudes Spatiales (France).


\bibliography{sample701}{}
\bibliographystyle{aasjournalv7.1}



\end{document}